\documentclass[journal]{IEEEtran}
\usepackage{amsmath,amsfonts,amsthm}
\usepackage{array}
\usepackage[caption=false,font=normalsize,labelfont=sf,textfont=sf]{subfig}
\usepackage{textcomp}
\usepackage{stfloats}
\usepackage{url}
\usepackage{verbatim}
\usepackage{graphicx}
\usepackage{xcolor}
\usepackage{nomencl}
\usepackage{etoolbox}
\usepackage{ifthen}
\usepackage{booktabs}
\usepackage[ruled]{algorithm2e}
\usepackage[normalem]{ulem}

\usepackage{multirow}
\usepackage{bm}
\usepackage{makecell}
\usepackage{enumitem}

\allowdisplaybreaks 
\usepackage[subtle,tracking=normal]{savetrees}

\usepackage{cite}

\def\BibTeX{{\rm B\kern-.05em{\sc i\kern-.025em b}\kern-.08em
    T\kern-.1667em\lower.7ex\hbox{E}\kern-.125emX}}
\usepackage{balance}

\theoremstyle{definition}

\newtheorem{theorem}{Theorem}

\newtheorem{prop}{Proposition}

\newtheorem{corollary}{Corollary}

\newtheorem{remark}{Remark}

\newcommand{\rev}[1]{\textcolor{blue}{#1}} 
\newcommand\Tstrut{\rule{0pt}{2.6ex}}         
\newcommand\Bstrut{\rule[-0.9ex]{0pt}{0pt}}   
\usepackage{url}
\usepackage{hyperref} 
\usepackage{xcolor}

\hypersetup{
    colorlinks=true, 
    linkcolor=blue, 
    filecolor=black, 
    urlcolor=blue, 
    citecolor=blue, 
    linkbordercolor={1 1 1}, 
    pdfborder={0 0 0} 
}
\begin{document}

\bstctlcite{IEEEexample:BSTcontrol}
\title{Locational Energy Storage Bid Bounds \\for Facilitating Social Welfare Convergence}


%

\author{\IEEEauthorblockN{Ning Qi, \textit{Member, IEEE}}, \IEEEauthorblockN{Bolun Xu, \textit{Member, IEEE}}

\thanks{Manuscript created 25 February, 2025; revised 22 May, 2025; accepted 11 June, 2025. This work was partly supported by the Department of Energy, Office of Electricity, Advanced Grid Modeling Program under contract DE-AC02-05CH11231 and partly supported by the National Science Foundation under award ECCS-2239046. Paper No. TEMPR-00048-2025. (\textit{Corresponding author}: Bolun Xu.)

Ning Qi and Bolun Xu are with the Department of Earth and Environmental Engineering, Columbia University, New York, NY
10027 USA (e-mail: \{nq2176, bx2177\}@columbia.edu). }
}

\markboth{IEEE TRANSACTIONS ON Energy Markets, Policy and Regulation,~Vol.~X, No.~X, XX June~2025}
{How to Use the IEEEtran \LaTeX \ Templates}

\maketitle


\begin{abstract}

This paper proposes a novel method to generate bid bounds that can serve as offer caps for energy storage in electricity markets to help reduce system costs and regulate potential market power exercises. We derive the bid bounds based on a tractable multi-period economic dispatch chance-constrained formulation that systematically incorporates the uncertainty and risk preference of the system operator. The key analytical results verify that the bounds effectively cap storage bids across all uncertainty scenarios with a guaranteed confidence level. We show that bid bounds decrease as the state of charge increases but rise with greater netload uncertainty and risk preference. We test the effectiveness of the proposed pricing mechanism based on the 8-bus ISO-NE test system, including agent-based storage bidding models. Simulation results demonstrate that the proposed bid bounds effectively align storage bids with the social welfare objective and outperform existing deterministic bid bounds. Under 30\% renewable capacity and 20\% storage capacity, the bid bounds contribute to an average reduction of 0.17\% in system cost, while increasing storage profit by an average of 10.16\% across various system uncertainty scenarios and bidding strategies. These benefits scale up with increased storage economic withholding and storage capacity.

\end{abstract}
\begin{IEEEkeywords}
Energy storage, locational bid bounds, chance-constrained optimization, market power, market design
\end{IEEEkeywords}

\section{Introduction}\label{Introduction}

Surging deployments of energy storage are introducing new challenges in regulating market power and facilitating social welfare convergence. As of March 2025, the capacity of battery energy storage in the California Independent System Operator (CAISO) has exceeded 11.3 GW and is projected to reach 50 GW by 2045~\cite{CAISORE}, with most storage units conducting price arbitrage in wholesale markets~\cite{zheng2023energy} while simultaneously submitting charge and discharge bids~\cite{williams2022electricity}. Market offerings of energy storage critically depend on future opportunities, which are difficult to quantify or benchmark~\cite{baker2023transferable}, fundamentally differing from thermal generators, whose market offers are based on fuel costs~\cite{kirschen2018fundamentals}. Hence, current market practices primarily rely on storage participants generating strategic bids, with a limited understanding of how these bids would impact system economics and exercise market power~\cite{xu2024truthful}. A recent study into historical energy storage bids in CAISO shows considerable practices of economic withholding and the resulting economic inefficiencies~\cite{ma2025comparative}. 

The challenge of monitoring energy storage market offers lies in calculating future opportunities caused by limited energy capacity~\cite{harvey2001market}. In day-ahead markets, storage may choose to withhold capacity to arbitrage in the more volatile real-time markets~\cite{qin2023role}, while in real-time markets, storage may withhold capacity in anticipation of future price spikes~\cite{ebrahimian2018price,yang2021real}. Consequently, inaccurate prediction of future uncertainty may lead to excessively high or low bids, resulting in market inefficiency. This has also been evidenced by storage practices observed in CAISO~\cite{ma2025comparative}. On the other hand, storage can exercise market power by conducting economic withholding, but identifying these intentions is extremely difficult due to the inability to distinguish from economic withholding for capturing legitimate future opportunities~\cite{xu2024truthful}. Hence, it is essential for system operators to develop innovative regulatory approaches to mitigate storage market power and facilitate storage bidding consistent with social welfare convergence.

This paper proposes a novel approach to imposing bid bounds on storage offers, providing system operators with a preventive measure to regulate market power and enhance market efficiency. The bounds dynamically depend on future system conditions, uncertainties, risk preference, and storage physical characteristics. Our contributions are as follows:
\begin{enumerate}
    \item \textit{Chance-Constrained Bounds:} 
        We propose a novel chance-constrained framework that provides locational storage bid bounds with a system cost minimization objective. The bounds are derived from a chance-constrained multi-period economic dispatch problem incorporating uncertainties from netload, as well as the risk preference of system operators.
    \item \textit{Theoretical Pricing Analysis:} 
We provide a theoretical analysis of key characteristics of the proposed bounds to establish a robust understanding of market intuitions. We prove that the bounds cap truthful bids with a confidence level and the cleared storage bids should be bounded by the risk-aware locational marginal price (LMP). We also show that the bid bounds decrease monotonically with the state of charge (SoC) and increase monotonically with netload uncertainty and risk preference.
\item \textit{Simulation Analysis:} 
        We test the proposed bid bound and benchmark it against a deterministic bound using an agent-based market simulation with varying storage bid strategies and system uncertainty scenarios on the modified 8-zone ISO-NE test system with multiple renewables and storage units. The proposed bid bound outperforms the existing deterministic bounds in reducing system costs and ensuring storage profits, and these benefits scale up with increased storage economic withholding and storage capacity.
\end{enumerate}


We organize the remainder of the paper as follows. 
Section~\ref{LR} summarizes the previous works on market design of energy storage and pricing of uncertainty. Section~\ref{PR}~provides problem formulation and preliminaries of chance-constrained pricing framework. Section~\ref{MR}~presents the theoretical pricing analysis. Section~\ref{Case Study}~describes case studies to verify the theoretical results. 
Finally, section~\ref{Conclusion}~concludes this paper.

\begin{figure}[t]
\setlength{\abovecaptionskip}{-0.1cm}  
\setlength{\belowcaptionskip}{-0.1cm} 
    \centering
\includegraphics[ width=0.95\columnwidth]
{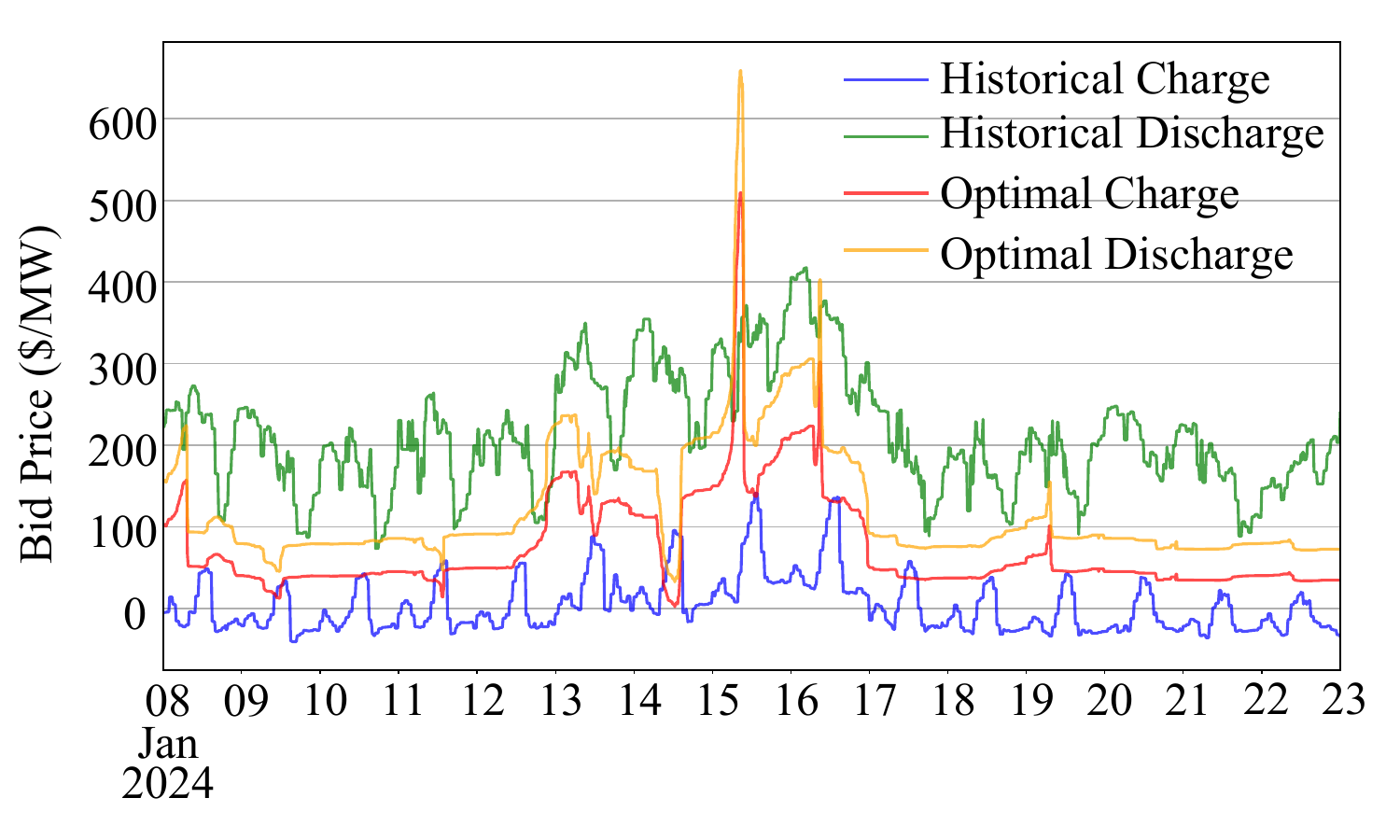}
    \caption{Comparison of hourly weighted average bids and  optimal hindsight bids during January 2024 from CAISO~\cite{ma2025comparative}.}
\label{OptimalHistoricalBids}
\end{figure}

\section{Backgrounds and Literature Review}\label{LR}

\subsection{Energy Storage Strategic Bidding}

Facilitated by FERC Order 841, U.S. power system operators now permit energy storage systems to submit charge (demand) and discharge (generation) bids to energy markets, recognizing that these bids must account for future opportunities and uncertainties~\cite{xu2024truthful}. For storage participants, developing optimal bidding strategies requires considering the physical characteristics of the storage, uncertainties in future market prices, and opportunities from participating in multiple markets. This complexity results in a challenging problem, as electricity prices are highly volatile and do not adhere to standardized process models.

Researchers have explored a diverse range of approaches to strategically operate energy storage considering physical and opportunity characteristics, including model predictive control~\cite{arnold2011model,yi2025perturbed}, stochastic optimization~\cite{krishnamurthy2017energy,fleten2007stochastic}, bi-level optimization~\cite{wang2017look}, robust optimization~\cite{wu2025energy}, reinforcement learning~\cite{wang2018energy,li2024temporal}, and model-based learning techniques~\cite{baker2023transferable, sang2022electricity}. However, these storage bidding strategies are typically designed from the perspective of individual storage participants rather than that of the system operators. They rely heavily on proprietary information, including price forecasts and uncertainty models, which poses challenges for power system operators in effectively monitoring and regulating these bids. 

In practice, profit-oriented strategic bidding by storage participants may compromise social welfare and reduce market efficiency. Take CAISO as an example, Figure~\ref{OptimalHistoricalBids} compares publicly available average bid data with hindsight optimal bids generated using an analytical dynamic programming approach based on historically cleared market prices~\cite{ma2025comparative}. These hindsight optimal bids represent the truthful marginal costs of storage given perfect information. The comparison reveals that actual discharge bids are significantly higher (and charge bids significantly lower) than the hindsight optimal bids, indicating strategic economic withholding by storage operators. Although the hindsight bids do not consider uncertainties or specific unit and locational factors, the substantial gap between historical and optimal bids highlights significant opportunities to enhance market efficiency.

\subsection{Market Efficiency Managements of Energy Storage}
Measures to enhance market efficiency in energy storage participation can be categorized into two main approaches: 1) deriving default bids to enable system operators to manage the dispatch of energy storage directly, and 2) mitigating market power in storage bids to prevent storage operators from exerting undue influence on market prices through strategic bidding behavior. However, both strategies are still in their early stages of development. For instance, CAISO recently proposes a default bid approach for 4-hr batteries~\cite{CAISODEB}, setting the default bid as the $4^{th}$ highest day-ahead LMP plus physical costs. Yet, this approach is overly conservative as it overlooks real-time uncertainties~\cite{xu2020operational} and the dependency of opportunity value on storage state-of-charge~\cite{zheng2023energy}. Consequently, CAISO limits enforcement of this default bid only in cases when local market power exercise has been evidently identified.

Direct adoption of existing storage bidding strategies for system-level storage management poses significant challenges to ensure reliability and efficiency. First, system operators require approaches that are both reliable and interpretable, which excludes most learning-based methods lacking transparency and cannot enforce constraints in exact. On the other hand, model-based methods that systematically incorporate uncertainty, such as stochastic dual dynamic programming~\cite{steeger2014optimal,garcia2025long,liu2025analytical,papavasiliou2017application}, often entail significant computational complexity, making them impractical when scaled to multiple networked storage units.

Recent studies have incorporated uncertainties~\cite{dvorkin2019chance,roald2017chance,xiao2023integrated} and temporal dependencies~\cite{hogan2020electricity,zhao2019multi} in market pricing primarily focused on ramping and reserve management. Yet, batteries typically require a 12- to 24-hour planning horizon to address daily price fluctuations between peak and valley periods~\cite{chen2021pricing}, while longer duration storage like hydropower may require months. These extended horizons exceed those typically associated with ramping or reserve operations. Additionally, most storage bids are submitted to wholesale energy markets, system operators need to maintain a neutral stance in market clearings while uphold reliability standards further complicates the integration of these models.

Detecting market manipulation through storage operations remains challenging due to the lack of an opportunity cost baseline~\cite{contereras2017energy}. Most studies in this area have been descriptive and have yet to produce definitive conclusions when applied to realistic operations under uncertainty. For example, recent research suggests that storage discharge bids should not exceed the highest daily price~\cite{zhou2024energy} and that storage operators should avoid withholding energy to manipulate market prices~\cite{wu2024market}. However, these recommendations are grounded in deterministic frameworks and fail to account for scenarios where storage operators use proprietary models incorporating price uncertainties to design their bids. To this end, the locational bid bounds proposed in this work serve as a preventive measure to mitigate potential market power abuse and bid inefficiencies, while still allowing storage participants the flexibility to develop their bid strategies within the established limits.

\section{Problem Formulation and Preliminaries}\label{PR}
Our objective is to derive probabilistic bounds for locational and unit-specific energy storage market offers. To do this, we begin by formulating an Oracle baseline economic dispatch problem that represents the optimal scenario for storage dispatch. Recognizing the uncertainties in demand and renewable generation, we then introduce chance constraints into the economic dispatch framework and develop a deterministic convex reformulation to establish robust bid bounds.



\subsection{Oracle Economic Dispatch}

The \textbf{Oracle economic dispatch problem (OED)} assumes a multi-period dispatch with perfectly forecasted demand and renewable profiles. While OED is not achievable in practice, it provides a baseline for our later analysis. OED is formulated as follows. 
\begin{subequations}\label{ED}
\begin{align}
&\min\ \sum_{t\in\mathcal{T}}[\sum_{i\in\mathcal{G}}{{{C}_{i}}( {{g}_{i\text{,}t}})}+\sum_{s\in\mathcal{S}}M_{s}(p_{s\text{,}t}+b_{s\text{,}t})]\label{obj}\\
&\text{s.t. }\forall i\in\mathcal{G}\text{, } \forall s\in\mathcal{S}\text{, } \forall l\in\mathcal{L}\text{, }\forall t\in\mathcal{T}  \nonumber \\
& \sum_{i\in\mathcal{G}}{g}_{i\text{,}t}+\sum_{s\in\mathcal{S}}(p_{s\text{,}t}-b_{s\text{,}t})=\sum_{n\in\mathcal{N}}{d}_{n\text{,}t}\text{: }{{\lambda }_{t}} \label{pb}\\
&\mid \sum_{n\in\mathcal{N}}\pi_{l-n}(\sum_{i\in\mathcal{N}_{n}}{{g}_{i\text{,}t}}+\sum_{s\in\mathcal{N}_{n}}({p}_{s\text{,}t}-{b}_{s\text{,}t})-{d}_{n\text{,}t})\mid\le\overline{F}_{l}\text{: }\underline\omega_{l\text{,}t}\text{,}\overline\omega_{l\text{,}t} \label{power-flow}\\
&\underline{G}_{i}\le {{g}_{i\text{,}t}}\le \overline{G}_{i}-r_{i\text{,}t}\label{rgulbound}\\
&\sum_{i\in\mathcal{G}} r_{i\text{,}t}\geq\rho\sum_{n\in\mathcal{N}}{d}_{n\text{,}t}\label{rs}\\
&-RD_{i}\le {{g}_{i\text{,}t}}-{{g}_{i\text{,}t-1}}\le RU_{i}\label{rgramp}\\
&0\le b_{s\text{,}t}\le \overline{P}_{s}\label{pcbound}\\
&0\le p_{s\text{,}t}\le \overline{P}_{s} \label{pdbound}\\
&b_{s\text{,}t}\perp p_{s\text{,}t} \label{st}\\
&\underline{E}_s \leq {e}_{s\text{,}t}\leq \overline{E}_s \label{rsulbound}\\
 &{e}_{s\text{,}t}-{e}_{s\text{,}t-1}=-p_{s\text{,}t}/{\eta}_{s}+b_{s\text{,}t}{{\eta }_{s}}\text{: }{{\theta }_{s\text{,}t}}\label{SoC}
\end{align}
\end{subequations}
\noindent where $\mathcal{G}$, $\mathcal{S}$, $\mathcal{T}$, $\mathcal{N}$, and $\mathcal{L}$ denote the sets of conventional generators, storages, time periods, nodes, and lines, respectively, and the subscripts $i$, $s$, $t$, $n$, $l$ correspond to the elements within these sets. ${C}_{i}$ and ${M}_{s}$ denote the production cost function of conventional generator and degradation cost of storage [\$/MWh]. ${{d}_{n\text{,}t}}$ denotes the netload (load minus renewable generation) [MWh]. $\overline{F}_{l}$ denotes the power limit of transmission line normalized per time step [MWh]. $\pi_{l-n}$ denotes the power transfer distribution factor from node $n$ to line $l$. $\rho$ defines the reserve capacity ratio of the conventional generator (e.g., $\rho=10\%$). $\overline{P}_{s}$\text{, }$\overline{E}_{s}$ and $\underline{E}_{s}$ denote the power capacity of storage, normalized per time step [MWh] and maximum and minimum SoC of storage [MWh]. $\eta_{s}$ denotes the one-way efficiency of storage. $\overline{G}_{i}$ and $\underline{G}_{i}$ denote the maximum and minimum power output of conventional generator, normalized per time step [MWh]. $\overline{RU}_{i}$ and $\underline{RD}_{i}$ denote the ramp-up and ramp-down limits of conventional generator, normalized per time step [MWh]. ${g}_{i\text{,}t}$\text{, }$r_{i\text{,}t}$\text{; }$p_{s\text{,}t}$\text{, }$b_{s\text{,}t}$ and $e_{s\text{,}t}$ denote the decision variables for dispatched energy and reserve energy of conventional generator [MWh]; discharge energy, charge energy, and SoC of storage [MWh]. ${\lambda }_{t}$, $\underline{\omega}_{l\text{,}t}$, $\overline{\omega}_{l\text{,}t}$, and ${\theta}_{s\text{,}t}$ are dual variables of corresponding constraints. ${\lambda }_{t}$ and ${\theta}_{s\text{,}t}$ denote the energy price, and LMP is defined as: $\text{LMP}_{n\text{,}t}=\lambda_{t}-\sum\nolimits_{l}\pi_{l-n}(\overline{\omega}_{l\text{,}t}-\underline{\omega}_{l\text{,}t})$.  

The objective function~\eqref{obj} minimizes system cost of conventional generators and storages. Constraints~\eqref{pb} guarantee the power balance. Constraints~\eqref{power-flow} limit the transmission line power flow. By using DC-OPF model~\cite{fang2021ac}, $\pi=B_{l}\hat{\bm{B}}^{-1}$, $B_{l}$ is the admittance of line $l$, \rev{$\hat{B}^{-1}$} is the  pseudoinverse matrix of bus admittance. Constraints~\eqref{rgulbound} limit the power output of conventional generators. Constraints~\eqref{rs} guarantee the reserve capacity from conventional generators. Constraints~\eqref{rgramp} limit the ramp-up/down of conventional generators.  Charge and discharge power of storage are limited by~\eqref{pcbound}-\eqref{pdbound}. Constraints~\eqref{st} prevent simultaneous charging and discharging of storage and can be reformulated using binary variables, rending a mixed-integer linear programming~\cite{qin2024economic}. By substituting the binary variables with their solutions, we can obtain dual variables from the reduced convex model~\cite{zhong2021chance}. Storage SoC is limited by constraints~\eqref{rsulbound}. Constraints \eqref{SoC} define the relationship between the SoC and charge/discharge energy of storage. 

\begin{remark}\emph{Marginal value of storage opportunity.} The dual variable ${\theta }_{s\text{,}t}$ the  associated with \eqref{SoC} is the marginal opportunity value of energy stored at storage $s$ at the end of time period $t$. This is trivial to show as \eqref{SoC} is the only time-coupling constraint of the storage operation. In later sections, we will use ${\theta }_{s\text{,}t}$ as a main reference to analyze and develop storage bid bounds.
\end{remark}

\subsection{Single period dispatch and storage marginal cost}
Practical real-time economic dispatch considers either a single time period or a very short look-ahead primarily for ramp rate management. To simplify the mathematical presentation and solely focus on storage management, we consider the following \textbf{single-period economic dispatch problem (SED)} defined as follows:
\begin{subequations}\label{sed}
\begin{align}
    & \min\ \sum_{i\in\mathcal{G}}{{{C}_{i}}( {{g}_{i\text{,}t}})}+\sum_{s\in\mathcal{S}}(A_{s\text{,}t}p_{s\text{,}t} - B_{s\text{,}t}b_{s\text{,}t}) \label{sed:obj}\\
    &\text{s.t. \eqref{pb}--\eqref{rgramp}} \nonumber\\
    &0\le b_{s\text{,}t}\le \min\{\overline{P}_{s}\text{,}\;(\overline{E}_{s}-e_{s\text{,}t-1})/\eta_s\}\label{sed:pcbound}\\
    &0\le p_{s\text{,}t}\le \min\{\overline{P}_{s}\text{,}\;(e_{s\text{,}t-1}-\underline{E}_{s})\eta_s\}\label{sed:pdbound}
\end{align} 
\end{subequations}
where $A_{s\text{,}t}$ and $B_{s\text{,}t}$ denote the storage discharge and charge marginal costs. Note $e_{s\text{,}t-1}$ is treated as a parameter instead of a variable as SED only optimizes time step $t$.

Given perfect foresight of net load over a future \textit{T}-period horizon, we calculate the optimal hindsight marginal cost of storage using Lagrangian relaxation, in which the time-coupling constraints~\eqref{SoC} are relaxed. The resulting optimal hindsight marginal cost shown in~\eqref{defaultbid}, includes both physical costs $M_s$ and opportunity costs $\theta_{s\text{,}t}$.
\begin{subequations}\label{defaultbid}
\begin{align}
A_{s\text{,}t}&=  \partial(M_{s}p_{s\text{,}t}+\theta_{s\text{,}t}p_{s\text{,}t}/{\eta_{s}})/{\partial p_{s\text{,}t}} =M_{s}+\theta_{s\text{,}t}/{\eta_{s}}\label{dischargebid}\\
 B_{s\text{,}t}&=  -\partial(M_{s}b_{s\text{,}t}-\theta_{s\text{,}t}b_{s\text{,}t}{\eta_{s}})/{\partial b_{s\text{,}t}} =\theta_{s\text{,}t}{\eta_{s}}-M_{s}\label{chargebid}
\end{align}
\end{subequations}
Note that due to uncertainties in netload and inter-temporal constraints~\eqref{SoC}, problem~\eqref{ED} is unsolvable in practice, and the opportunity cost $\theta_{s,t}$ cannot be known. 

\subsection{Chance-Constrained Multi-Period Economic Dispatch}
We now extend \eqref{ED} to incorporate netload uncertainties using a \textbf{chance-constrained economic dispatch formulation (CED)}. The chance-constrained approach is employed due to its favorable balance between probabilistic interpretability and computational tractability. Compared to stochastic programming~\cite{kazempour2018stochastic,wong2007pricing}, the chance-constrained method explicitly incorporates probabilistic metrics (i.e., mean and standard deviation) and allows operators to define clear risk tolerances through adjustable confidence levels. And the chance-constrained approach is typically less conservative than robust optimization~\cite{velloso2019two,zugno2015robust}, as it allows system operators to explicitly control the degree of conservatism by adjusting the confidence level. By doing so, operators can directly manage the likelihood of constraint violations, ensuring system reliability. Moreover, chance-constrained optimization problems can often be exactly reformulated into deterministic equivalents, enabling efficient solution methods at scale~\cite{roald2017chance,dvorkin2019chance}. 

The CED provides probabilistic bounds on the storage opportunity costs, which serve as a base for designing market offer caps. The CED is formulated in~\eqref{MCCED}. For short-duration storage (e.g., batteries), which typically cycles within one day, the time horizon of the CED can be naturally set to one day. In contrast, for long-duration storage such as hydropower, the relevant time horizon should be extended to several months. Given the inherent inaccuracies or inadequacies of long-term probabilistic forecasts, the system operator can leverage historical scenarios combined with continuously updated real-time probabilistic forecasts~\cite{qi2025long,garcia2025long} to derive the bid bounds.
\begin{subequations}\label{MCCED}
\begin{align}
&\min\ \sum_{t\in\mathcal{T}}[\sum_{i\in\mathcal{G}}{{{C}_{i}}( {{g}_{i\text{,}t}})}+\sum_{s\in\mathcal{S}}M_{s}(p_{s\text{,}t}+b_{s\text{,}t})]\label{obj1}\\
&\text{s.t. }\forall i\in\mathcal{G}\text{, } \forall s\in\mathcal{S}\text{, } \forall l\in\mathcal{L}\text{, }\forall t\in\mathcal{T}  \nonumber \\
& \mathbb P\big(\sum_{i\in\mathcal{G}}{g}_{i\text{,}t}+\sum_{s\in\mathcal{S}}(p_{s\text{,}t}-b_{s\text{,}t})\geq\sum_{n\in\mathcal{N}}{d}_{n\text{,}t})\geq1-\epsilon\text{: }{\hat{\lambda }_{t}} \label{pb1}\\
&\mathbb P\big(\mid \sum_{n\in\mathcal{N}}\pi_{l-n}(\sum_{i\in\mathcal{N}_{n}}{{g}_{i\text{,}t}}+\sum_{s\in\mathcal{N}_{n}}({p}_{s\text{,}t}-{b}_{s\text{,}t})-{d}_{n\text{,}t})\mid\leq\overline{F}_{l}\big)\nonumber\\
&\geq1-\epsilon\text{: }\hat{\underline\omega}_{l\text{,}t}\text{,}\hat{\overline\omega}_{l\text{,}t} \label{power-flow1}\\
&\underline{G}_{i}\le {{g}_{i\text{,}t}}\le \overline{G}_{i}-r_{i\text{,}t}\text{: }\hat{\underline{\nu }}_{i\text{,}t}\text{,}\hat{\overline{\nu }}_{i\text{,}t}\label{rgulbound1}\\
&\mathbb P\big(\sum_{i\in\mathcal{G}} r_{i\text{,}t}\geq\rho\sum_{n\in\mathcal{N}}{d}_{n\text{,}t}\big)\geq1-\epsilon\label{rs1}\\
&-RD_{i}\le {{g}_{i\text{,}t}}-{{g}_{i\text{,}t-1}}\le RU_{i}\text{: }\hat{\underline{\kappa }}_{i\text{,}t}\text{,}\hat{\overline{\kappa }}_{i\text{,}t}\label{rgramp1}\\
&0\le b_{s\text{,}t}\le \overline{P}_{s}\text{: }\hat{\underline{\alpha}}_{s\text{,}t}\text{,}\hat{\overline{\alpha }}_{s\text{,}t} \label{pcbound1}\\
&0\le p_{s\text{,}t}\le \overline{P}_{s}\text{: }\hat{\underline{\beta }}_{s\text{,}t}\text{,}\hat{\overline{\beta }}_{s\text{,}t} \label{pdbound1}\\
&b_{s\text{,}t}\perp p_{s\text{,}t} \label{st1}\\
&\underline{E}_s \leq {e}_{s\text{,}t}\leq \overline{E}_s \text{: }\hat{\underline{\iota}}_{s\text{,}t}\text{,}\hat{\overline{\iota}}_{s\text{,}t}\label{rsulbound1}\\
&{e}_{s\text{,}t}-{e}_{s\text{,}t-1}=-p_{s\text{,}t}/{\eta}_{s}+b_{s\text{,}t}{{\eta }_{s}}\text{: }{\hat{\theta }_{s\text{,}t}}\label{SoC1}
\end{align}
\end{subequations}
\noindent where $\mathbb{P}$ denotes the probability function for chance-constraints. $\epsilon$ denotes the probability level of chance-constraints, e.g., $\epsilon$=5\%. $\lambda_{t}$, $\theta_{s\text{,}t}$, $\underline{\omega}_{l\text{,}t}$, $\overline{\omega}_{l\text{,}t}$, $\underline{\nu}_{i\text{,}t}$, $\overline{\nu}_{i\text{,}t}$, $\underline{\kappa}_{i\text{,}t}$, $\overline{\kappa}_{i\text{,}t}$, $\underline{\alpha}_{s\text{,}t}$, $\overline{\alpha}_{s\text{,}t}$, $\underline{\beta}_{s\text{,}t}$, $\overline{\beta}_{s\text{,}t}$, $\underline{\iota}_{s\text{,}t}$, and $\overline{\iota}_{s\text{,}t}$ are dual variables of corresponding constraints under CED framework. $\hat\lambda_{t}$ and $\hat{\theta}_{s\text{,}t}$ denotes the energy price and opportunity cost of storage derived from CED framework. And risk-aware LMP is defined as: $\hat{\text{LMP}}_{n\text{,}t}=\hat\lambda_{t}-\sum\nolimits_{l}\pi_{l-n}(\hat{\overline{\omega}}_{l\text{,}t}-\hat{\underline{\omega}}_{l\text{,}t})$.  

Compared to the deterministic framework, constraints \eqref{pb}, \eqref{power-flow}, and \eqref{rs} are changed into chance-constraints \eqref{pb1}, \eqref{power-flow1}, and \eqref{rs1}, which ensure that these constraints are simultaneously satisfied with a $1-\epsilon$ confidence level. The objective function for cost minimization and all other constraints remain unchanged.

\subsection{Problem Reformulation}

Chance-constrained optimization requires a deterministic reformulation to solve the problem and derive prices from dual variables. Given an assumed probability distribution, with accurately estimated statistical parameters (mean $\mu_{t}$ and standard deviation $\sigma_{t}$), chance-constraints \eqref{pb1}, \eqref{power-flow1}, and \eqref{rs1} admit a deterministic reformulation in \eqref{dr}. 
\begin{subequations}\label{dr}
    \begin{align}
& \sum_{i\in\mathcal{G}}{g}_{i\text{,}t}+\sum_{s\in\mathcal{S}}(p_{s\text{,}t}-b_{s\text{,}t})\geq\sum_{n\in\mathcal{N}}(\mu_{n\text{,}t}+F^{-1}(1-\epsilon)\sigma_{n\text{,}t})\\
&  \sum_{n\in\mathcal{N}}\pi_{l-n}(\sum_{i\in\mathcal{N}_{n}}{{g}_{i\text{,}t}}+\sum_{s\in\mathcal{N}_{n}}({p}_{s\text{,}t}-{b}_{s\text{,}t})-\mu_{n\text{,}t}-\\
&F^{-1}(1-\epsilon)\sigma_{n\text{,}t}\geq-\overline{F}_{l}\text{, }\sum_{n\in\mathcal{N}}\pi_{l-n}(\sum_{i\in\mathcal{N}_{n}}{{g}_{i\text{,}t}}+\sum_{s\in\mathcal{N}_{n}}({p}_{s\text{,}t}-{b}_{s\text{,}t})\nonumber\\
&-\mu_{n\text{,}t}+F^{-1}(1-\epsilon)\sigma_{n\text{,}t}) \le\overline{F}_{l}\nonumber\\
&  \sum_{i\in\mathcal{G}} r_{i\text{,}t}\geq\rho\sum_{n\in\mathcal{N}}(\mu_{n\text{,}t}+F^{-1}(1-\epsilon)\sigma_{n\text{,}t})
\end{align}
\end{subequations}
\noindent where  $F^{-1}(\cdot)$ is the normalized inverse cumulative distribution function (CDF) of netload.

\begin{remark}\emph{Uncertainty models.}
To clearly illustrate how the inverse CDF can be obtained under different uncertainty assumptions, we summarize the following cases.
\begin{enumerate}
\item For uncertainty with assumed distribution (e.g., normal distribution), the inverse CDF is directly obtained from the known distribution, e.g., $F^{-1}(1-\epsilon)=\Phi^{-1}(1-\epsilon)$ for normal distribution.
\item For uncertainty with ambiguous or partial information, we can obtain the robust approximation of inverse CDF from generalizations of the Cantelli’s inequality~\cite{qi2023}, as summarized in Table~\ref{approximation}.
\item For uncertainties represented by discrete historical observations, a parametric "Versatile Distribution" proposed by~\cite{versatile} can be adopted, where parameters (a,b,c) are learned via Maximum Likelihood Estimation, leading to a closed-form inverse CDF expression: $F^{-1}(1-\epsilon\mid a\text{,}b\text{,}c)=c-\ln \left((1-\epsilon)^{-1 / b}-1\right)/{a}$.
\end{enumerate}
In practice, the inverse CDF under complex uncertainty structures (e.g., with spatio-temporal correlations) can also be estimated using methods such as Sample Average Approximation~\cite{vrakopoulou2017chance} or Distributionally Robust Optimization~\cite{tan2024robust,zugno2015robust}.
\end{remark}
\begin{table}[t]
  \centering
  \caption{Robust Approximation of Normalized Inverse Cumulative Distribution with Ambiguous Information}
  \setlength{\tabcolsep}{0.2mm}{
      \begin{tabular}{c c c}
    \toprule
    Type \& Shape & ${{F}^{-1}}(1-\epsilon)$ & $\epsilon$ \\
    \midrule
    No Assumption (NA) & $\sqrt{(1-\epsilon )/\epsilon }$ & $0<\epsilon \le 1$ \\\specialrule{0em}{0.5em}{0em}
    \multirow{2}[0]{*}{ Symmetric Distribution (S)} & $\sqrt{1/2\epsilon }$ & $0<\epsilon \le 1/2$ \\\specialrule{0em}{0.5em}{0em}
          & 0     & $1/2<\epsilon \le 1$ \\\specialrule{0em}{0.5em}{0em}
    \multirow{2}[0]{*}{ Unimodal Distribution (U)} & $\sqrt{(4-9\epsilon )/9\epsilon }$ & $0<\epsilon \le 1/6$ \\\specialrule{0em}{0.5em}{0em}
          & $\sqrt{(3-3\epsilon )/(1+3\epsilon )}$ & $1/6<\epsilon \le 1$ \\\specialrule{0em}{0.5em}{0em}
    \multirow{3}[0]{*}{{ Symmetric \& Unimodal Distribution (SU)}} & $\sqrt{2/9\epsilon }$ & $0<\epsilon \le 1/6$ \\\specialrule{0em}{0.5em}{0em}
          & $\sqrt{3}(1-2\epsilon )$ & $1/6<\epsilon \le 1/2$ \\\specialrule{0em}{0.5em}{0em}
          & 0     & $1/2<\epsilon \le 1$ \\\specialrule{0em}{0.5em}{0em}
    \bottomrule
    \end{tabular}%
    }
  \label{approximation}%
\end{table}%

\section{Main Results}\label{MR}

In this section, we derive the locational bid bounds for storage participants. We then present a theoretical analysis to demonstrate the monotonicity of bid bounds with respect to SoC, uncertainty and risk preference.

\subsection{Locational Bid Bounds}\label{OER}

Our main result shows the storage opportunity value dual $\hat\theta_{s\text{,}t}$ from the CED problem serves as a probabilistic bound of the storage marginal cost from the OED problem.

\begin{theorem}\label{tm}  \emph{Locational storage bid bounds.}
The chance-constrained locational storage bid bounds are formulated as:
\begin{subequations}\label{bidbounds}
\begin{align}
&\mathbb P(A_{s\text{,}t}\leq \overline{A}_{s\text{,}t} )\geq 1-\epsilon\text{, } \mathbb P(B_{s\text{,}t}\leq \overline{B}_{s\text{,}t})\geq 1-\epsilon \label{bidboundrelation}\\
&\overline{A}_{s\text{,}t}=M_{s}+\max(\hat\theta_{s\text{,}t})/{\eta_{s}}\text{, } \overline{B}_{s\text{,}t}=\min(\hat\theta_{s\text{,}t}){\eta_{s}}-M_{s}\label{bidboundformulation}
\end{align}
\end{subequations}
\end{theorem}

\noindent where $\overline{A}_{s\text{,}t}$ and $\overline{B}_{s\text{,}t}$ represent the discharge and charge bid bounds, respectively, derived from the chance-constrained framework as presented in~\eqref{bidboundformulation}. These bid bounds serve as upper limits for the truthful marginal costs as shown in~\eqref{bidboundrelation}.

Theorem~\ref{tm} demonstrates that for cleared storage units, the bids submitted by storage participants should be limited by the proposed bounds with a $1-\epsilon$ confidence level. However, as shown in~\eqref{max}, if power or SoC constraints are binding, the storage unit is not cleared, allowing the bids to exceed these bounds. Hence, the system operator can distinguish it from exercising market power based on the proposed bounds and storage states. We defer the complete proof to Appendix~\ref{appendix1}.

Due to the limited knowledge of system uncertainties, storage struggles to bid efficiently, making it difficult to capture legitimate future opportunities. Overbidding or underbidding can reduce storage profits and social welfare. To address this, storage bid bounds can benchmark storage market power and effectively adjust bids to facilitate social welfare convergence.

\begin{corollary}\label{c1}
\emph{Anticipated storage bid bounds.} The locational storage bid bounds equal the risk-aware LMP.
\begin{align}\label{LMP-bidbounds}
\overline{A}_{s\text{,}t}=\max(\hat{\text{LMP}}_{m\text{,}t})\text{, }\overline{B}_{s\text{,}t}=\min(\hat{\text{LMP}}_{m\text{,}t})    
\end{align} 
\end{corollary}  
Corollary~\ref{c1} shows that the system operator can anticipate storage bid bounds based on the projected risk-aware LMP. Under perfect foresight, truthful storage bids coincide with LMPs, as storage would optimally charge at the minimum LMP and discharge at the maximum LMP. However, incorporating uncertainty elevates these bid bounds to risk-aware LMP levels. This explains why storage participants rationally increase their bids under future uncertainty, highlighting the importance and necessity of explicitly incorporating risk into market design. This result aligns with previous work~\cite{qin2024economic} under a price-taker profit maximization framework, which shows that the upper bound of storage bids is limited by peak energy price.

\subsection{SoC Monotonicity of Storage Bid Bounds}
We further show that the storage bid bounds are a convex function of storage SoC. Hence, they serve as bounds for SoC-dependent bids~\cite{zheng2023energy}, enabling the efficient integration of SoC-dependent bids into the market clearing. To demonstrate this, the following proposition proves that the storage bid bounds monotonically decrease with SoC.

\begin{remark}\emph{Second-order differentiability and piece-wise linear approximations.} We assume that the generator cost function is convex second-order differentiable for the following theoretical analysis. Hence, given a quadratic or super-quadratic cost function $C_{i}$, $\partial^{2}C_{i}(g_{i\text{,}t})/\partial g_{i\text{,}t}^{2}\geq0$. For application to piecewise-linear/quadratic cost functions, we can approximate using discrete second-order derivative calculations:
\begin{align}
\frac{\partial^{2}C_{i}(g_{i\text{,}t})}{\partial g_{i\text{,}t}^{2}} \approx \frac{C_i(g_{i\text{,}t} + \Delta g)+C_i(g_{i\text{,}t} - \Delta g)-2C_i(g_{i\text{,}t})}{\Delta g^2}
\end{align}
where $\Delta g$ is a sufficiently small step, then according to the convex definition (monotonic increasing cost functions), the second-order derivative approximation is always non-negative.
\end{remark}

\begin{prop}\label{p2}
\emph{SoC-dependent storage bid bounds.}
   Given a monotonically increasing and quadratic or super-quadratic cost function~$C_{i}$, we have $\partial \overline{A}_{s\text{,}t}/{\partial {{e}_{s\text{,}t-1}}}\le 0$, $\partial\overline{B}_{s\text{,}t}/{\partial {{e}_{s\text{,}t-1}}}\le 0$. 
\end{prop}

These results fit the diminishing storage value, i.e., the marginal value decreases with storage SoC levels.  Proposition~\ref{p2} also aligns with prior studies that show the storage opportunity value function is convex, but were derived using a price-taker profit maximization framework~\cite{xu2020operational, zheng2023energy}. In contrast, we derive proposition~\ref{p2} using a social welfare maximization framework. These results provide convex bounds for SoC-dependent storage bids, demonstrating that a convex cost function can guide both system operator dispatch and the bids submitted by storage participants. We defer the complete proof to Appendix~\ref{appendix2}.

\subsection{Uncertainty Monotonicity of Storage Bid Bounds}
We further show that storage bid bounds increase with system uncertainty and risk preference of system operators. This is a significant difference compared to conventional generators, whose bids should always be based on fixed fuel cost curves, independent of uncertainty and risk preference.

\begin{prop}\label{p3}\emph{Storage bid bounds scaling with system uncertainty.}
Given a quadratic or super-quadratic function \( C_{i} \), we have \( \partial \overline{A}_{s\text{,}t}/{\partial \sigma_{n\text{,}t}} \geq 0 \), \( \partial\overline{B}_{s\text{,}t}/{\partial \sigma_{n\text{,}t}} \geq 0 \).
\end{prop}

Note that the uncertainties lie not only in the renewables and load but also in the look-ahead window of the dispatch. Proposition~\ref{p3} demonstrates that with a quadratic or super-quadratic function of \( C_{i} \), the storage bid bounds will increase with higher non-anticipativity either from higher penetration of uncertain resources (e.g., renewables, flexible load) or a longer look-ahead window. This result can also confirm that the bid bounds incorporating future uncertainty ($\sigma_{t}>0$) should be greater than those without uncertainty consideration ($\sigma_{t}=0$). We defer the complete proof to Appendix~\ref{appendix2}.

Proposition~\ref{p3} also aligns with prior studies~\cite{xu2020operational, qin2024economic} that have shown the storage bids scale with price uncertainty in price-taker profit-maximization objectives. Yet, we derive this result considering netload uncertainty from a social welfare maximization perspective.

\begin{prop}\label{p4}\emph{Storage bid bounds scaling with risk preference.}
Given a quadratic or super-quadratic function \( C_{i} \), we have \( \partial \overline{A}_{s\text{,}t}/{\partial \epsilon} \leq 0 \), \( \partial\overline{B}_{s\text{,}t}/{\partial \epsilon} \leq 0 \).
\end{prop}

Proposition~\ref{p4} shows that the system operator can adjust the bid bounds based on their risk preference. However, when $\epsilon$ is too small, storage bid bounds become excessively large, reducing their effectiveness in limiting market power. Conversely, when $\epsilon$ is too large, storage bid bounds become too conservative, failing to recover the truthful storage cost, which diminishes both storage profitability and social welfare. Therefore, the system operator should choose a trade-off value for $\epsilon$ in practice. We defer the complete proof to Appendix~\ref{appendix2}.



\section{Numerical Case Study}\label{Case Study}

\subsection{Agent-based Experiment Setups}
We use a modified agent-based simulation framework with strategic storage participants~\cite{qin2023role} based on the ISO-NE system~\cite{krishnamurthy20158} to demonstrate the effectiveness of proposed bid bounds in system costs and storage profits. It should be noted that the proposed framework and corresponding results are transferable and not limited to this specific test system, due to the generality of the adopted agent-based simulation approach. The test system consists of 8 nodes, 12 lines, 76 thermal generators with a total installed capacity of 23.1 GW, and an average load of 13 GW. 

Our experiment has the following procedures:
\begin{enumerate}
    \item Select 10 representative day-ahead (DA) netload scenarios and 100 Monte Carlo samples of real-time (RT) realizations for each DA scenario based on a pre-set netload uncertainty, hence 1000 scenarios for each trial.
    \item Perform DA unit commitment to derive the day-ahead price for each scenario (see \cite{qin2023role} for the DA unit commitment formulation).
    \item For each of the ten DA scenarios, repeat for the following procedure for each of the 100 real-time scenarios:
    \begin{enumerate}
        \item Generate storage economic withholding bids based on the day-ahead price and assumed price derivations $\sigma$, please refer to Appendix~\ref{appendix3} for detailed bid generation formulations. 
        \item Generate the proposed bid bounds as~\eqref{bidbounds}.
        \item Perform RT economic dispatch using the SED formulation in \eqref{sed} with economic withholding bids submitted by storage participants. The SED runs sequentially for 24 hours for each real-time scenario.
        \item Repeat SED with capped storage economic withholding bids using bid bounds.
    \end{enumerate}
    \item Record the averaged system cost and storage profit results across scenarios and samples of steps 3-c) and 3-d).
\end{enumerate}
We also repeated the experiment trials with different energy storage withholding and system uncertainty.

For clarity of presentation, we linearly map the price standard deviation $\sigma$ (ranging from 0 to 50 \$/MWh) onto an economic capacity withholding scale of 1–5, reflecting that higher price uncertainty leads to increased economic capacity withholding in storage bids design~\cite{xu2020operational}. The baseline uncertainty scenarios are derived from the Elia dataset~\cite{Elia-data}, with the netload uncertainty scale represented by scaling the baseline netload standard deviation $\sigma$ from 1 to 3 times its original value. Unless otherwise specified, renewables are set at 30\% of maximum load capacity, with storage configured at 20\% of maximum load capacity and 4-hour duration. Initial SoC, efficiency, and marginal cost of storage are set to be 0.5, 95\%, and \$10/MWh. The probabilistic level is set to be 5\%. 

The optimization is coded in MATLAB and solved by Gurobi 11.0 solver. The programming environment is Intel Core i9-13900HX @ 2.30GHz with RAM 16 GB\footnote{The code and data used in this study are available at: \url{https://github.com/thuqining/Storage_Pricing_for_Social_Welfare_Maximization.git}}.

\begin{figure}[t]
  \begin{center}  \includegraphics[width=1\columnwidth]{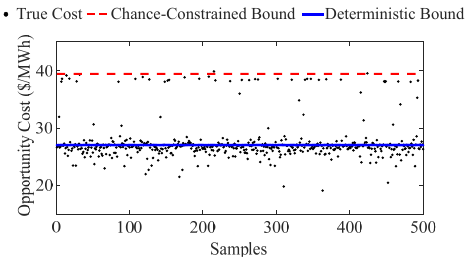}
    \caption{Truthful opportunity cost and corresponding bounds.}\label{figuretheorem}
  \end{center} 
\end{figure}

 \begin{figure}[t]
  \begin{center}  \includegraphics[width=0.9\columnwidth]{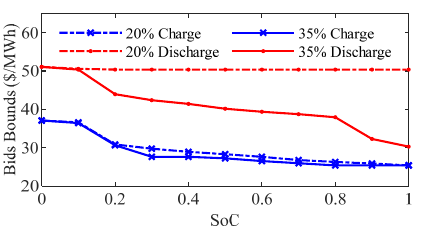}
    \caption{Storage discharge and charge bid bounds variations with SoC under 20\% and 35\% storage capacity scenarios.}\label{figsoc}
\end{center}
\end{figure}

\subsection{Analysis on Bid Bounds Effectiveness and Dependency}
We first demonstrate some key characteristics of the bid bound, which directly validates the analytical results presented in the previous sections.
\subsubsection{Bound effectiveness}
Figure~\ref{figuretheorem} compares the truthful maximum opportunity costs from 500 Monte Carlo samples of netload uncertainty realizations with the proposed chance-constrained opportunity cost bound (with confidence level $\epsilon=5\%$) and the deterministic opportunity cost bound motivated by the current CAISO default bids (the 4\textsuperscript{th} highest day-ahead LMP). The proposed opportunity cost bounds effectively cap the true opportunity cost of storage with a guaranteed confidence level. Hence, the bid bounds which consist of fixed physical cost and opportunity cost bound can reliably cap the truthful bids submitted by storage participants. This result verifies the Theorem~\ref{tm}. While the deterministic opportunity cost bound lies approximately at the median level across different scenarios, it is more likely to underestimate the truthful opportunity costs, potentially leading to reduced storage profits.


\subsubsection{SoC-Dependent Storage Bid Bounds}

Figure~\ref{figsoc} demonstrates that storage bid bounds decrease monotonically with SoC, confirming Proposition~\ref{p2}. Comparing the trends at 20\% and 35\% storage capacities, the 35\% case exhibits a significantly stronger dependency on SoC. This indicates that higher storage capacity has a more pronounced impact on system operations and marginal costs. In particular, the lower discharge bid bound at the 35\% case suggests that storage operations further drove down the system LMPs.


\subsubsection{Uncertainty Scaling of Storage Bid Bounds}

\begin{figure}[t]
  \begin{center}  \includegraphics[width=0.9\columnwidth]{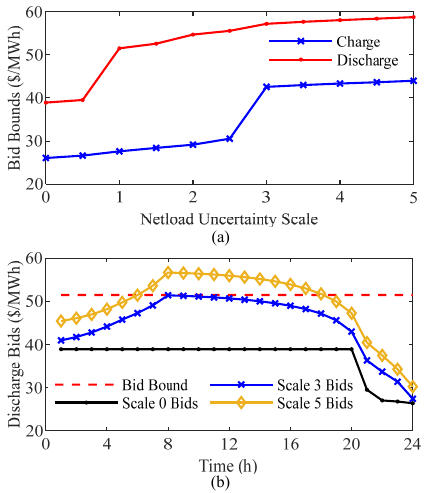}
    \caption{Storage bid bounds: (a) variations with netload uncertainty and (b) comparison to storage economic withholding bids.}\label{figuncertainty}
  \end{center}
\end{figure}

Figure~\ref{figuncertainty} (a) demonstrates that the storage bid bounds monotonically increase with uncertainty, which verifies the Proposition~\ref{p3}. The discharge and charge bid bounds increase by 50\% and 68\%, respectively, as uncertainty scales from 0 to 5.

The bid bounds can benchmark storage economic withholding behavior. 
As illustrated in Figure~\ref{figuncertainty} (b), the bids submitted by storage participants increase monotonically with the economic withholding scale.  Therefore, significant economic withholding results in excessively high discharge bids or low charge bids, thereby affecting storage clearing and ultimately impacting social welfare. The proposed bid bounds adjust inefficient storage bids by clipping any values that exceed these bounds. It is observed that bids with scale 0 and 3 economic withholding levels are reasonable, whereas bids with scale 5 economic withholding exceed the storage’s truthful marginal cost and are consequently capped by the bid bounds. Moreover, the bounds can help storage understand the system uncertainty level. Under the current netload uncertainty, the assumed price $\sigma$ is expected to be around 15 \$/MWh.


\begin{figure*}[t]
  \begin{center}  \includegraphics[width=0.95\linewidth]{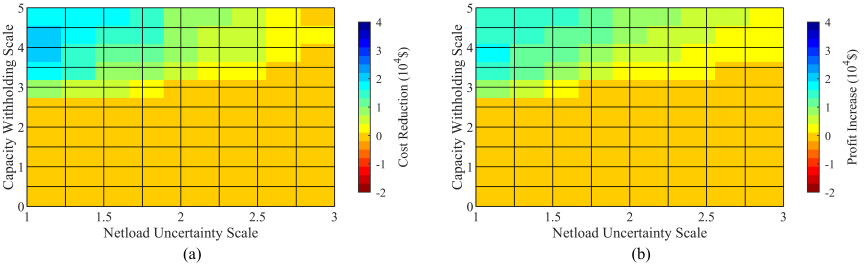}
    \caption{Difference after adding storage bounds under 20\% storage capacity and 30\% renewable capacity: (a) system cost reduction and (b) storage profit increase.}\label{figurecompare1}
  \end{center}
\end{figure*}

\begin{figure*}[t]
  \begin{center}  \includegraphics[width=0.95\linewidth]{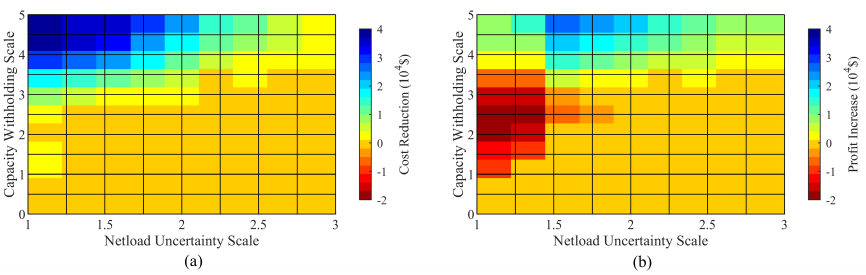}
    \caption{Difference after adding storage bounds 35\% storage capacity and 30\% renewable capacity: (a) system cost reduction and (b) storage profit increase. }\label{figurecompare2}
  \end{center}
\end{figure*}

\subsection{Agent-based System Operation Analysis}
We now show the agent-based simulation results.

\subsubsection{Impact on system cost and storage profits}
Figure~\ref{figurecompare1} compares the system cost and storage profit at 30\% renewable capacity and 20\% storage capacity.  The results indicate that the proposed bounds reduce system cost while increasing storage profits in scenarios of high storage withholding and low system uncertainty (upper-left regions of the figures). This improvement arises because the bid bound mitigates inefficient storage bids—those with excessively high withholding that unduly limits system availability. By capping such bids, storage availability is enhanced, leading to lower system costs and higher profits.

Figure~\ref{figurecompare2} offers a similar comparison under a scenario with higher storage capacity—30\% renewable capacity and 35\% storage capacity. Compared to Figure~\ref{figurecompare1}(a), the 35\% case yields more significant system cost savings, indicating that the bid bound's contribution increases with greater storage shares. Conversely, Figure~\ref{figurecompare2}(b) shows that in low uncertainty conditions with moderate withholding levels (mid-left region), the bid bound reduces storage profit. In this region, storage units achieve higher prices and profits through optimal withholding, capping these bids lowers profits while cutting system costs. Notably, the storage profit reduction accounts for only 10–20\% of the system cost savings. Therefore, the system operator can leverage these cost savings to establish compensation or incentive mechanisms for storage participants who voluntarily comply with the bid bounds. Such an incentive-based framework can help maintain fairness and encourage compliance, ensuring both individual profitability and system-level efficiency. At higher withholding levels (greater than 4), the bounds again prove beneficial by curbing inefficient bids that would otherwise undermine profitability, mirroring the trends observed in Figure~\ref{figurecompare1}.

\begin{table*}[!t]
  \centering
  \caption{Impact of Storage and Renewable Capacity on Economic Performance and Improvement Under Low and High Uncertainty}
  \begin{tabular}{cc|cccc|cccc}
    \toprule
    \multirow{2}{*}{Renewable } & \multirow{2}{*}{Storage } & \multicolumn{4}{c|}{Low Uncertainty} & \multicolumn{4}{c}{High Uncertainty} \\
    \cmidrule{3-10}
     &  & \multicolumn{2}{c}{System Cost ($10^6$\$(\%))} & \multicolumn{2}{c|}{Storage Profit ($10^5$\$(\%))} & \multicolumn{2}{c}{System Cost ($10^6$\$(\%))} & \multicolumn{2}{c}{Storage Profit ($10^5$\$(\%))} \\
     &  & AEW & MEW & AEW & MEW & AEW & MEW & AEW & MEW \\
    \midrule
    \multirow{3}{*}{30\%} 
      & 20\% & 7.63(-0.17) & 7.65(-0.23) & 1.04(10.16) & 0.92(14.48) & 7.80(-0.06) & 7.81(-0.09) & 0.94(4.96) & 0.92(7.72) \\
      & 35\% & 7.57(-0.21) & 7.60(-0.48) & 1.48(0.19)  & 1.29(12.24) & 7.74(-0.09) & 7.78(-0.18) & 1.40(4.77) & 1.37(10.69) \\
      & 50\% & 7.48(-0.20) & 7.52(-0.40) & 2.03(0.90)  & 1.90(6.83)  & 7.64(-0.07) & 7.68(-0.12) & 2.00(1.47) & 2.00(3.89) \\
    \midrule
    \multirow{3}{*}{50\%} 
      & 20\% & 7.50(-0.11) & 7.52(-0.18) & 0.91(6.63)  & 0.78(11.68) & 7.72(-0.02) & 7.72(-0.04) & 0.85(1.31) & 0.83(2.71) \\
      & 35\% & 7.42(-0.10) & 7.45(-0.23) & 1.44(-1.10) & 1.30(2.21)  & 7.63(-0.03) & 7.66(-0.06) & 1.40(1.13) & 1.38(2.46) \\
      & 50\% & 7.37(-0.08) & 7.40(-0.26) & 1.77(-1.51) & 1.69(2.44)  & 7.57(-0.05) & 7.59(-0.07) & 1.80(0.62) & 1.85(1.89) \\
    \bottomrule
  \end{tabular}
  \label{merged-table}
\end{table*}

\subsubsection{Result sensitivity to storage and renewable capacity} 
Table~\ref{merged-table} provides more comprehensive results of the economic performance and its improvement with different storage and renewable capacities under low system uncertainty (averaged over 1-1.5 scale) and high system uncertainty (averaged over 1.75-3 scale), with average economic withholding (AEW) cases averaged over withholding scale from 0 to 5, and the maximum economic withholding (MEW) corresponding to the withholding scale of 5. The result shows under all scenarios, storage bid bounds can reliably reduce the system cost, with the highest reduction case close to 0.5\%. On the other hand, the bid bound improves the most profit at low storage capacity levels, for it helps to modulate less efficient bids from storage that overly withhold capacity.  

An increase in renewable capacity leads to lower energy prices, thereby diminishing the effectiveness of bid bounds. Notably, in high-renewable scenarios, storage profit is sacrificed by an average of 1.10–1.15\%. Under high uncertainty scenarios, both system cost and storage profit are elevated relative to low uncertainty scenarios. However, the effectiveness of bid bounds is reduced, since it becomes more rational to withhold higher capacity as indicated by the higher bid bounds.

\subsubsection{Result sensitivity to uncertainty model and risk preference} 
We evaluate the performance of the proposed pricing mechanism under different uncertainty models as shown in Table~\ref{uncertaintymodel}. It is important to emphasize that the assumed uncertainty model significantly affects both the deterministic reformulation of the chance-constraints and the resulting bid bounds. Although the mean and standard deviation remain consistent across different models, variations in the inverse CDF lead to considerable differences in the performance. The empirical model serves as the benchmark, in which empirical quantiles directly obtained from historical data are used for the deterministic reformulation. 

The empirical model generates the lowest bid bounds and achieves the highest social welfare improvement compared to the others. The versatile distribution demonstrates the best performance in fitting uncertainty, as it can capture skewness and multimodal characteristics, with results closely aligning with the empirical model. The robust approximation results in the least social welfare improvement due to its excessively high bid bounds, particularly in high uncertainty scenarios. Compared to the empirical model, the versatile distribution demonstrates relatively better performance, with only a 0.1\%-0.2\% reduction. This indicates that system operators can guarantee acceptable performance using versatile distribution models.

Furthermore, we compare the performance of the Gaussian model under different risk preferences by varying $\epsilon$. Table~\ref{epsilon} shows that as $\epsilon$ decreases, bid bounds increase while social welfare improvement declines, which verifies the Proposition~\ref{p4}. Moreover, under low uncertainty, the bid bounds and the associated performance are more sensitive to $\epsilon$. For a high $\epsilon$ setting, the bid bounds may prevent storage from recovering its truthful cost in extreme scenarios, whereas for a low $\epsilon$ setting, excessively high bid bounds reduce their effectiveness in limiting strategic economic withholding behavior. Especially, when $\epsilon$ is set to 15\%, system cost and storage profit decrease by an average of 0.20\% and 13.77\%, respectively.  

We also tested deterministic bid bounds, whose performance closely resembles the proposed bid bounds with $\epsilon=50\%$, although our proposed bounds are SoC-dependent. Under the low-uncertainty scenario, deterministic bounds yield a slightly lower cost reduction of 0.18\% compared to the chance-constrained bounds. More significantly, under the high uncertainty scenario, deterministic bounds increase system costs by 0.01\% due to underestimated storage opportunity costs, as illustrated in Figure~\ref{figuretheorem}. Consequently, storage profits are compromised by approximately 20\% across both low and high uncertainty scenarios. These results suggest that system operators should adopt chance-constrained bid bounds with $\epsilon$ set around 5\% to 10\%.

\begin{table}[!ht]
\centering
\caption{Comparison of Economic Performance and Improvement under Different Uncertainty Models}
  \setlength{\tabcolsep}{0.1mm}{
\begin{tabular}{c c c c c c}
\toprule
\makecell{Uncertainty\\Scale} & 
\makecell{Model} & 
\makecell{System Cost\\($10^6\$$)} & 
\makecell{Cost\\Reduction (\%)} & 
\makecell{Storage Profit\\($10^5\$$)} & 
\makecell{Profit\\Increase (\%)} \\
\midrule
\multirow{4}{*}{1.0} 
  & Empirical & \multirow{4}{*}{\centering 7.60} & -0.14 & \multirow{4}{*}{\centering 1.06} & 8.96 \\
  & Versatile &                          & -0.15 &                            & 9.71 \\
  & Gaussian  &                          & -0.19 &                            & 11.21 \\
  & Robust    &                          & -0.16 &                            & 10.41 \\
\midrule
\multirow{4}{*}{3.0}
  & Empirical & \multirow{4}{*}{\centering 7.93} &  0.05 & \multirow{4}{*}{\centering 0.90} & 5.24 \\
  & Versatile &                          & -0.04 &                            & 4.52 \\
  & Gaussian  &                          & -0.01 &                            & 3.09 \\
  & Robust    &                          & -0.00 &                            & 1.75 \\
\bottomrule
\end{tabular}\label{uncertaintymodel}
}
\end{table}

\begin{table}[!ht]
  \centering
  \caption{Comparison of Economic Performance and Improvement under Different Risk Preference}
  \setlength{\tabcolsep}{0.5mm}{
    \begin{tabular}{c c c c c c}
    \toprule
    \makecell{Uncertainty\\Scale} & 
    \makecell{\(\epsilon\)\\(\%)} & 
    \makecell{System Cost\\($10^6\$$)} & 
    \makecell{Cost\\Reduction (\%)} & 
    \makecell{Storage Profit\\($10^5\$$)} & 
    \makecell{Profit\\Increase (\%)} \\
    \midrule
    \multirow{4}{*}{1.0} 
      & 15 & \multirow{4}{*}{7.60} & -0.20 & \multirow{4}{*}{1.06} & -13.77 \\
      & 10 &                    & -0.19 &                    & 11.45 \\
      & 5  &                    & -0.19 &                    & 11.21 \\
      & 1  &                    & -0.17 &                    & 10.52 \\
    \midrule
    \multirow{4}{*}{3.0} 
      & 15 & \multirow{4}{*}{7.93} & -0.11 & \multirow{4}{*}{0.90} & 8.96 \\
      & 10 &                    & -0.10 &                    & 8.92 \\
      & 5  &                    & -0.01 &                    & 3.09 \\
      & 1  &                    & -0.00 &                    & 1.99 \\
    \bottomrule
    \end{tabular}
  }
  \label{epsilon}
\end{table}

\begin{table}[!ht]
  \centering
  \caption{Comparison of Economic Performance and Improvement under Scarcity Scenarios}
  \setlength{\tabcolsep}{0.3mm}{
    \begin{tabular}{ccccc}
    \toprule
    \makecell{Generation\\Curtailment Ratio} & \makecell{Max Real-Time\\ Price ($\$$)} & \makecell{Withholding\\ Scale} & \makecell{System Cost\\($10^6\$$)} & \makecell{Storage Profit\\($10^5\$$)} \\
    \midrule
    \multirow{2}[1]{*}{0} & 52.02 & 5     & 7.43 (-0.03) & 0.32 (2.59) \\
          & 57.58 & 10    & 7.51 (-0.04) & 0.24 (14.86) \\
    \multirow{2}[0]{*}{20\%} & 78.31 & 5     & 9.74 (-0.07) & 1.73 (8.15) \\
          & 90.32 & 10    & 9.75 (-0.23) & 1.26 (23.36) \\
    \multirow{2}[1]{*}{40\%} & 218.63 & 5     & 15.28 (-0.21) & 7.48 (4.02) \\
          & 234.92 & 10    & 15.36 (-1.27) & 7.11 (25.41) \\
    \bottomrule
    \end{tabular}
  }
  \label{scenario}
\end{table}

\subsubsection{Result sensitivity to scarcity scenarios}

We further investigate the effectiveness of our proposed bid bounds under scarcity scenarios characterized by extremely high prices and storage bids. Specifically, we simulate fault scenarios by curtailing generation to trigger high-cost generation dispatch or loss-of-load costs. And we evaluate individual real-time scenario performance instead of averaged Monte Carlo results. Table~\ref{scenario} shows that as real-time prices increase, both system costs and storage profits rise due to electricity scarcity, and the effectiveness of the proposed bid bounds improves. Additionally, the proposed bid bounds also improve with higher prices. Furthermore, as the economic withholding scale increases, system costs grow while storage profits decrease, and the effectiveness of the bid bounds is enhanced under higher economic withholding scenarios. This is consistent with our earlier findings under normal scenarios. These results indicate that the proposed bid bounds perform particularly well under scarcity scenarios.

\subsubsection{Computational efficiency and scalability} 
Table~\ref{time} shows the computing time for storage bid bounds calculation increases exponentially with the number of integrated storage units. When the number of storage units exceeds 5000, the problem takes over an hour to solve, making it impractical for real-world implementation. To address this problem, we employ a robust relaxation~\cite{relaxation} to avoid the use of binary variables, significantly enhancing computational performance. The computing time increases linearly with the number of storage units, requiring only 102.51 s for 10000 units.

\begin{table}[!ht]
  \centering
  \caption{Comparison of Computational Performance under Different Storage Number and
   Relaxation Condition}
    \setlength{\tabcolsep}{1.5mm}{
    \begin{tabular}{cccccc}
    \toprule
    \multicolumn{1}{c}{\multirow{2}[4]{*}{\makecell{Storage \\Number}}} & \multicolumn{2}{c}{CPU Time} & \multicolumn{1}{c}{\multirow{2}[4]{*}{\makecell{Storage \\Number}}} & \multicolumn{2}{c}{CPU Time} \\
\cmidrule{2-3}\cmidrule{5-6}          & \multicolumn{1}{l}{\makecell{Without\\ Relaxation}} & \multicolumn{1}{l}{\makecell{With\\ Relaxation}} &       & \multicolumn{1}{l}{\makecell{Without\\ Relaxation}} & \multicolumn{1}{l}{\makecell{With\\ Relaxation}} \\
    \midrule
    5     & 0.58 s  & 0.19 s  & 500   & 3.36 s  & 0.89 s  \Bstrut\\
    10    & 0.96 s  & 0.22 s  & 1000  & 317.99 s  & 2.04 s  \Tstrut\Bstrut\\
    50    & 1.86 s  & 0.24 s  & 5000  & $>$1h & 39.92 s  \Tstrut\Bstrut\\
    100   & 3.26 s  & 0.30 s  & 10000 & $>$1h & 102.51 s  \Tstrut\\
    \bottomrule
    \end{tabular}}
  \label{time}%
\end{table}%

\section{Conclusion and Discussion}\label{Conclusion}
We proposed a novel approach to generate bounds for capping energy storage market offers to help reduce system operating costs and regulate storage profits. These bounds are unit-location specific and generated using a tractable chance-constrained economic dispatch formulation that internalizes the netload uncertainty and the system operator's risk preference. We provide theoretical proof showing that the bid bounds cap truthful storage bids and has strong dependency with SoC, system uncertainty and risk preference. Agent-based numerical simulations based on the 8-zone ISO-NE test system verify our theoretical findings and show the proposed approach outperforms the existing deterministic bid bounds and can reliably reduce system cost and regulate storage profit, especially mitigating extreme withholding cases that also improve storage profits.


Our work addresses the pressing need for new regulatory approaches to manage energy storage market offers in electricity markets, while acknowledging that storage participants have valid causes for conducting economic withholding, which is sensitive to price volatility and uncertainty. Our approach enables operators to remain neutral, fostering competition among strategic storage participants, while capping offers to prevent excessive withholding that could compromise system efficiency. Additionally, the bounds can be tuned within a chance-constrained framework based on risk preferences and uncertainty models, allowing power system operators to update bids in line with their uncertainty profiles without directly influencing market-clearing outcomes.


The proposed framework provides a practical solution to managing the surge of storage participants in many electricity markets such as in California. As outlined in our motivation, storage participants are overly withholding their availability and evidently contributing to price spikes during periods when storage was not planned to discharge. Our approach considers a generalized economic dispatch model and is broadly transferable to various electricity markets. Our bound framework can be implemented as a simplification of the real market clearing models that ensures computation efficiency while offering insights into facilitating social welfare convergence as power systems scale up renewable and energy storage deployments. 


\section*{Acknowledgment}
The authors thank Sergio Duenas Melendez, Storage Sector Manager at the California Independent System Operator (CAISO), for valuable discussions that informed the practical context of this work.

\bibliographystyle{IEEEtran}
\bibliography{IEEEabrv,CCOP}

\appendix

\subsection{Proof of Theorem 1} \label{appendix1}
We provide the Karush-Kuhn-Tucker (KKT) conditions of CED~\eqref{MCCED} in~\eqref{KKT} for the following theoretical analysis.
\begin{subequations}
\begin{align}
 \dfrac{\partial L}{\partial g_{i\text{,}t}}&=\dfrac{\partial {{{C}_{i}}( {{g}_{i\text{,}t}})}}{\partial g_{i\text{,}t}}+\sum\nolimits_{l}\pi_{l-n}(\hat{\overline{\omega}}_{l\text{,}t}-\hat{\underline{\omega}}_{l\text{,}t})\label{lg2}\\
&-\hat{\lambda}_{t}-\hat{\underline\nu}_{i\text{,}t}+\hat{\overline\nu}_{i\text{,}t}-\hat{\underline{\kappa}}_{i\text{,}t}+\hat{\overline{\kappa}}_{i\text{,}t}=0\text{, }i\in\mathcal{G}_{n}\nonumber\\
 \dfrac{\partial L}{\partial b_{s\text{,}t}}&={M}_{s}-\sum\nolimits_{l}\pi_{l-m}(\hat{\overline{\omega}}_{l\text{,}t}-\hat{\underline{\omega}}_{l\text{,}t})-\hat{\underline{\alpha}}_{s\text{,}t}\label{dc2}\\
&+\hat{\overline{\alpha}}_{s\text{,}t}+\hat{\lambda}_{t}+(-\hat{\theta }_{s\text{,}t}-\hat{\underline\iota}_{s\text{,}t}+\hat{\overline\iota}_{s\text{,}t}){{\eta }}_{s}=0\text{, }s\in\mathcal{S}_{m}\nonumber\\
 \dfrac{\partial L}{\partial p_{s\text{,}t}}&={M}_{s}+\sum\nolimits_{l}\pi_{l-m}(\hat{\overline{\omega}}_{l\text{,}t}-\hat{\underline{\omega}}_{l\text{,}t})-\hat{\underline{\beta }}_{s\text{,}t}\label{dd2}\\
&+\hat{\overline{\beta }}_{s\text{,}t}-\hat{\lambda}_{t}+(\hat{\theta }_{s\text{,}t}+\hat{\underline\iota}_{s\text{,}t}-\hat{\overline\iota}_{s\text{,}t})/{{\eta }_{s}}=0\text{, }s\in\mathcal{S}_{m}\nonumber\\
 \dfrac{\partial L}{\partial e_{s\text{,}t}}&={\hat{\theta }_{s\text{,}t}}-{\hat{\theta }_{s\text{,}t+1}}-\hat{\underline\iota}_{s\text{,}t}+\hat{\overline\iota}_{s\text{,}t}=0\text{, }s\in\mathcal{S}_{m}\label{de2}
\end{align}\label{KKT}
\end{subequations}

From~\eqref{defaultbid}, the bid bounds should include both physical cost and opportunity cost bounds, hence we first prove the opportunity cost bounds:
\begin{subequations}\label{results1}
\begin{align}
&\mathbb P(\min(\hat\theta_{s\text{,}t})\geq\min({\theta}_{s\text{,}t}))\geq 1-\epsilon \\
&\mathbb P(\max(\hat\theta_{s\text{,}t})\geq\max({\theta}_{s\text{,}t})\geq{\theta}_{s\text{,}t})\geq 1-\epsilon
\end{align}
\end{subequations}

From~\eqref{dc2}-\eqref{dd2}, we derive the linear relationships between opportunity costs and risk-aware LMPs under charge and discharge states in~\eqref{linear}.
\begin{subequations}\label{linear}
\begin{align}
  \hat\theta_{s\text{,}t}&=(\hat{\text{LMP}}_{m\text{,}t}+{M}_{s}-\hat{\underline{\alpha}}_{s\text{,}t}+\hat{\overline{\alpha}}_{s\text{,}t})/{{\eta }}_{s}-\hat{\underline\iota}_{s\text{,}t}+\hat{\overline\iota}_{s\text{,}t}\label{theta1}\\
 \hat\theta_{s\text{,}t}&=(\hat{\text{LMP}}_{m\text{,}t}-{M}_{s}+\hat{\underline{\beta}}_{s\text{,}t}-\hat{\overline{\beta}}_{s\text{,}t}){{\eta }}_{s}-\hat{\underline\iota}_{s\text{,}t}+\hat{\overline\iota}_{s\text{,}t}\label{theta2} 
\end{align}    
\end{subequations}

For the cleared unit under charge state, we have $\hat{\underline\alpha}_{s\text{,}t}=\hat{\underline\iota}_{s\text{,}t}=0$, while under discharge state, we have $\hat{\underline\beta}_{s\text{,}t}=\hat{\overline\iota}_{s\text{,}t}=0$. Given that all dual variables are non-negative, we derive the minimum bound of charge opportunity cost in~\eqref{thetamin} and the maximum bound of discharge opportunity cost in~\eqref{thetamax}.
\begin{subequations}\label{max}
\begin{align}
  \min(\hat\theta_{s\text{,}t})&=(\min(\hat{\text{LMP}}_{m\text{,}t})+{M}_{s})/{{\eta }}_{s}\label{thetamin}\\
 \max(\hat\theta_{s\text{,}t})&=(\max(\hat{\text{LMP}}_{m\text{,}t})-{M}_{s}){{\eta }}_{s}\label{thetamax} 
\end{align}    
\end{subequations}

Given that we have a marginal generator unit $i$ for each time slot, i.e., $g_{j\text{,}t}=\overline{G}_{j} \text{ or } \underline{G}_{j}\text{, } j\neq i$. Hence, the constraints~\eqref{rgulbound1} and~\eqref{rgramp1} for the marginal unit $i$ are not binding, we have $\hat{\underline\nu}_{i\text{,}t}=\hat{\overline\nu}_{i\text{,}t}=\hat{\underline\kappa}_{i\text{,}t}=\hat{\overline\kappa}_{i\text{,}t}=0$. Then, combining~\eqref{lg2} and~\eqref{pb1}, we have~\eqref{LMP-c}. $\hat{\text{LMP}}_{m\text{,}t}^{\text{c}}$ and $\hat{\text{LMP}}_{n\text{,}t}^{\text{c}}$ denote the congestion cost of the storage node and marginal generator node, respectively. ${d}^{+}_{n\text{,}t}$ denotes the $(1-\epsilon)\%$ quantile of netload distribution. Similarly, the deterministic LMP is formulated in~\eqref{LMP-d}.  
\begin{subequations}
\begin{align} \hat{\text{LMP}}_{m\text{,}t}&=\dfrac{\partial  C_{i} \big(\sum\limits_{n\in\mathcal{N}}{d}^{+}_{n\text{,}t}-\sum\limits_{s\in\mathcal{S}}(p_{s\text{,}t}-b_{s\text{,}t})-\sum\limits_{j\in\mathcal{G}\text{, }j\neq i}{g}_{j\text{,}t} \big)}{\partial g_{i\text{,}t}}\nonumber\\
&+\hat{\text{LMP}}_{m\text{,}t}^{\text{c}}-\hat{\text{LMP}}_{n\text{,}t}^{\text{c}}\label{LMP-c} \\
\text{LMP}_{m\text{,}t}&=\dfrac{\partial  C_{i} \big(\sum\limits_{n\in\mathcal{N}}{d}_{n\text{,}t}-\sum\limits_{s\in\mathcal{S}}(p_{s\text{,}t}-b_{s\text{,}t})-\sum\limits_{j\in\mathcal{G}\text{, }j\neq i}{g}_{j\text{,}t} \big)}{\partial g_{i\text{,}t}}\nonumber\\
&+\text{LMP}_{m\text{,}t}^{\text{c}}-\text{LMP}_{n\text{,}t}^{\text{c}}\label{LMP-d}
\end{align}       
\end{subequations}

Since ${d}^{+}_{n\text{,}t}$ is larger than any realization of ${d}_{n\text{,}t}$ with a $1-\epsilon$ confidence level, and congestion is more severe under the chance-constrained framework, we have~\eqref{proof1}. By substituting~\eqref{proof1} into~\eqref{max}, we have proved~\eqref{results1}. Hence, we can derive the bid bounds based on the opportunity cost bounds and have finished the proof.
\begin{align}\label{proof1}
&\mathbb P(\text{LMP}_{m\text{,}t}\geq\hat{\text{LMP}}_{m\text{,}t})\geq 1-\epsilon
\end{align}

\subsection{Proof of Proposition 1-3} \label{appendix2}
\subsubsection{SoC-dependent storage bid bounds}
\begin{proof}
By substituting~\eqref{pb} and \eqref{LMP-c} into~\eqref{LMP-bidbounds}, we have: 
\begin{align}
 \dfrac{\partial{\overline{A}_{s\text{,}t}}}{\partial{e_{s\text{,}t-1}}}&=\dfrac{\partial^{2} C_{i} \left( g_{i\text{,}t} \right)}{\partial^{2} g_{i\text{,}t}}\dfrac{\partial{{g }_{i\text{,}t}}}{\partial{p_{s\text{,}t}}}\dfrac{\partial{p_{s\text{,}t}}}{\partial{e_{s\text{,}t-1}}}\label{case1}\\
 &=-{\eta_{s}}\partial^{2}  C_{i} \left( g_{i\text{,}t} \right)/{\partial^{2} g_{i\text{,}t}}\leq 0\nonumber\\
 \dfrac{\partial{\overline{B}_{s\text{,}t}}}{\partial{e_{s\text{,}t-1}}}&=\dfrac{\partial^{2} C_{i} \left( g_{i\text{,}t}(\bm{\xi}_{t}) \right)}{\partial^{2} g_{i\text{,}t}}\dfrac{\partial{{g }_{i\text{,}t}}}{\partial{b_{s\text{,}t}}}\dfrac{\partial{b_{s\text{,}t}}}{\partial{e_{s\text{,}t-1}}}\label{case2}\\
&=-\partial^{2} C_{i} \left( g_{i\text{,}t}(\bm{\xi}_{t}) \right)/{\eta_{s}\partial^{2} g_{i\text{,}t}}\leq 0\nonumber
\end{align}
Hence, we have finished the proof.
\end{proof}
\subsubsection{Storage bid bounds scaling with system uncertainty}
\begin{proof}
From~\eqref{LMP-c} and quadratic or super-quadratic function \( C_{i} \), we have:
\begin{align}\label{uncertainty}
   \dfrac{\partial \hat{\text{LMP}}_{m\text{,}t} }{ \partial \sigma_{n\text{,}t}}= \dfrac{\partial^2  C_i\left( g_{i\text{,}t} \right) }{\partial g_{i\text{,}t} \partial \sigma_{n\text{,}t}} &=     \dfrac{\partial^2  C_i\left( g_{i\text{,}t} \right) }{\partial^2 g_{i\text{,}t}}F^{-1}(1-\epsilon)\geq 0
\end{align}
By substituting~\eqref{uncertainty} into~\eqref{LMP-bidbounds}, we have finished the proof.
\end{proof}
\subsubsection{Storage bid bounds scaling with risk preference}
\begin{proof}
From~\eqref{LMP-c} and quadratic or super-quadratic function \( C_{i} \), we have:
\begin{align}\label{risk}
  \dfrac{\partial \hat{\text{LMP}}_{m\text{,}t} }{ \partial \epsilon}= \dfrac{\partial^2  C_i\left( g_{i\text{,}t} \right) }{\partial g_{i\text{,}t} \partial \epsilon} &=-\dfrac{\partial^2  C_i\left( g_{i\text{,}t} \right) }{\partial^2 g_{i\text{,}t}}\dfrac{\partial F^{-1}(1-\epsilon)\sigma_{n\text{,}t}}{\partial\epsilon}\leq 0
\end{align}
By substituting~\eqref{risk} into~\eqref{LMP-bidbounds}, we have finished the proof.
\end{proof}

\subsection{Formulation of Storage Economic Withholding Bids} \label{appendix3}

\textbf{(1) Opportunity Value Function.} The storage profit maximization is formulated in~\eqref{value} to derive the storage opportunity value function. To handle the SoC dependencies in the storage
model and uncertainty in price, stochastic dynamic programming can be used to recursively update the value function. The storage opportunity value function is determined by the mean of real-time price (i.e., day-ahead price), and monotonically increases with the standard deviation of real-time price~\cite{qin2024economic}. Hence, storage can exercise more economic withholding with higher assumed price uncertainty.
\begin{align}
& Q_{s\text{,}t-1}({e}_{s\text{,}t-1})=\max\ \lambda_t\left(p_{s\text{,}t}-b_{s\text{,}t}\right)-M_{s} (p_{s\text{,}t}+b_{s\text{,}t})\label{value}\\
&\hspace{2.1cm}+V_{s\text{,}t}({e}_{s\text{,}t})\notag\\
& V_{s\text{,}t}({e}_{s\text{,}t-1})=\mathbb E(Q_{s\text{,}t-1}({e}_{s\text{,}t-1})\mid\lambda_t)\notag\\
& \text{s.t. \eqref{pcbound}--\eqref{SoC}}\nonumber 
\end{align}

\noindent where $V_{s\text{,}t}$ is the opportunity value of energy storage, hence value-to-go function in the stochastic dynamic programming. 

\textbf{(2) Storage Economic Withholding Bids.} Energy storage can generate charge and discharge bids as~\eqref{eq:bidcurve}. 
\begin{subequations}\label{eq:bidcurve}
\begin{align}
 &A_{s\text{,}t} =M_{s}+v_{s\text{,}t}\left(e_{s\text{,}t-1}-p_{s\text{,}t} / \eta_{s}\right)/{\eta_{s}} \\ 
 &B_{s\text{,}t} =\eta_{s} v_{s\text{,}t}\left(e_{s\text{,}t-1}+b_{s\text{,}t} \eta_{s}\right)-M_{s}
\end{align}
\end{subequations}

\noindent where $v_{s\text{,}t}$ is the subderivative of $V_{s\text{,}t}$.

\begin{IEEEbiography}[{\includegraphics[width=1in,height=1.25in]{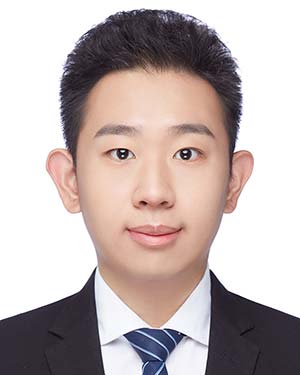}}]{Ning Qi}
    (Member, IEEE) was born in 1996. He received a B.S. degree in Electrical Engineering from Tianjin University, Tianjin, China, in 2018 and the Ph.D. degree in Electrical Engineering from Tsinghua University, Beijing, China, in 2023. He is currently a postdoctoral research scientist in Earth and Environmental Engineering at Columbia University. Before joining Columbia University, he was a visiting scholar at the Technical University of Denmark in 2022. He was a research associate in Electrical Engineering at Tsinghua University in 2024.
    His current research focuses on data-driven modeling, optimization under uncertainty, and market design for power systems with generalized energy storage.
\end{IEEEbiography}

\begin{IEEEbiography}[{\includegraphics[width=1in,height=1in]{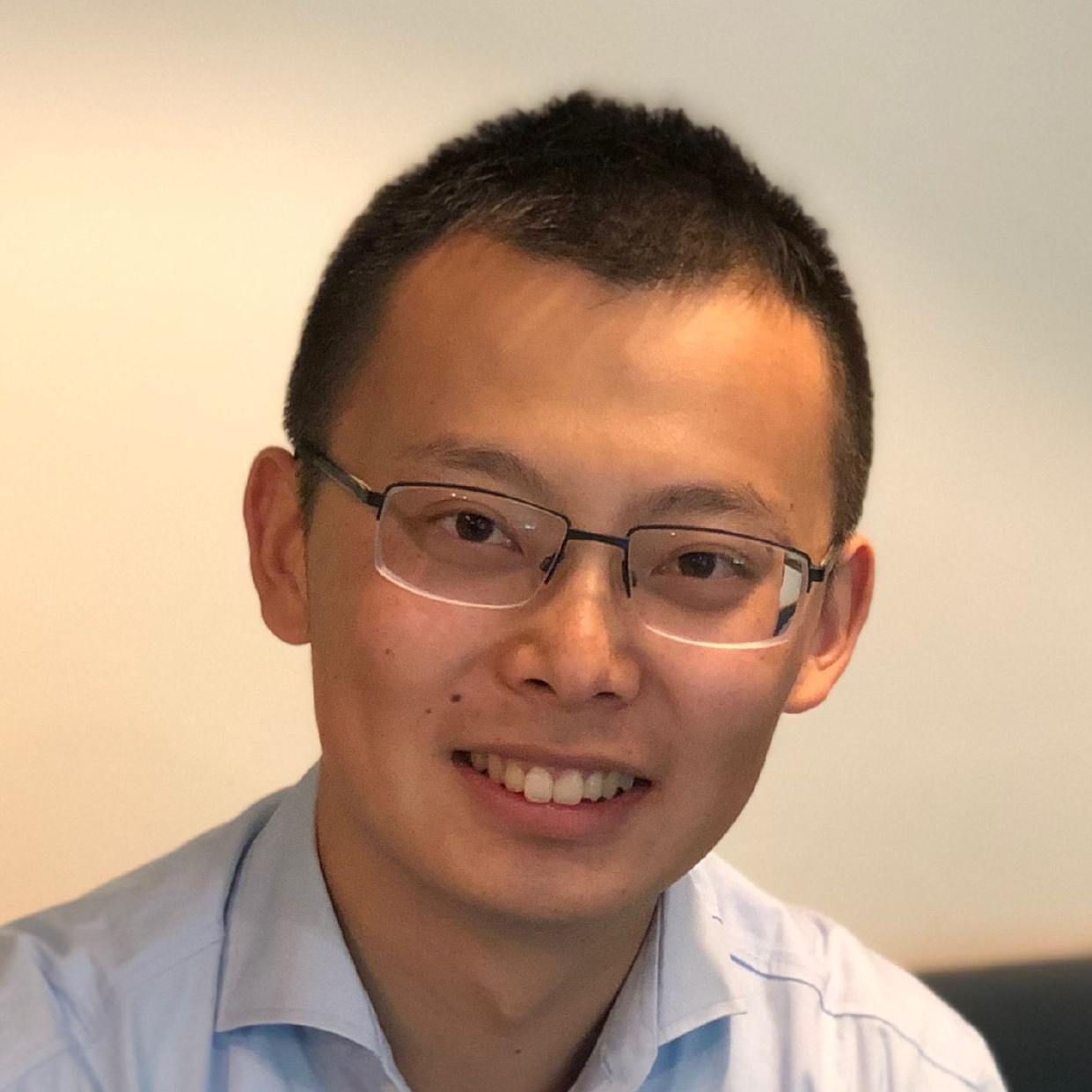}}]{Bolun Xu}
    (Member, IEEE) received the B.S. degree in electrical engineering from Shanghai Jiaotong University, Shanghai, China, in 2011, the M.Sc. degree in electrical engineering in the Swiss Federal Institute of Technology, Zurich, Switzerland, in 2014, and the Ph.D. degree in electrical engineering from the University of Washington, Seattle, WA, USA, in 2018. He is currently an Assistant Professor of Electrical Engineering with the Department of Earth and Environmental Engineering, Columbia University, New York, NY, USA. His research interests include energy storage, power system optimization, and power system economics.
\end{IEEEbiography}

\end{document}